\documentclass[journal]{IEEEtran}
\usepackage[numbers,compress]{natbib}

\usepackage{algorithmic}
\usepackage{amsmath}
\usepackage[ruled]{algorithm2e}
\usepackage{tabularx}
\usepackage{lipsum}

\usepackage{graphicx}

\begin{document}\sloppy
	
\title{A Novel Adaptation Method for HTTP Streaming of VBR Videos over Mobile Networks}

\author{Hung. T Le$^1$,	Hai N. Nguyen$^2$, Nam Pham Ngoc$^2$, Anh T. Pham$^1$, Truong Cong Thang$^1$ 

	\thanks{Hung T. Le, Anh T. Pham and Truong Cong Thang are with the University of Aizu, Aizu-Wakamatsu, Japan, email: \{d8162102, pham, thang\}@u-aizu.ac.jp.}
	\thanks{Hai N. Nguyen and Nam Pham Ngoc are with Hanoi University of Science and Technology, Hanoi, Vietnam, email: nam.phamngoc@hust.edu.vn.}
}

\maketitle

\bibliographystyle{IEEEtranN}

\begin{abstract}
Recently, HTTP streaming has become very popular for delivering video over the Internet. For adaptivity, a provider should generate multiple versions of a video as well as the related metadata. Various adaptation methods have been proposed to support a streaming client in coping with strong bandwidth variations. However, most of existing methods target at constant bitrate (CBR) videos only. In this paper, we present a new method for quality adaptation in on-demand streaming of variable bitrate (VBR) videos. To cope with strong variations of VBR bitrate, we use a local average bitrate as the representative bitrate of a version. A buffer-based algorithm is then proposed to conservatively adapt video quality. Through experiments, we show that our method can provide quality stability as well as buffer stability even under very strong variations of bandwidth and video bitrates.
\end{abstract}


\section{Introduction}
\label{intro}
Recently, video streaming has rapidly gained popularity over the Internet. It is predicted that global video traffic will reach 80 percent of total consumer Internet traffic in 2019 \cite{Cisco2015}. Besides, HTTP protocol has become a cost-effective solution thanks to the abundance of Web platform and broadband connections \cite{Begen2011_1,Thang2012}. Furthermore, for interoperability of HTTP streaming in the industry, ISO/IEC MPEG has developed ``Dynamic Adaptive Streaming over HTTP" (DASH) \cite{Stockhammer2011} as the first standard for video streaming over HTTP.

Due to the heterogeneity of communication networks nowadays, adaptivity is the principal requirement for any streaming clients. In DASH, multiple versions of an original video as well as related metadata (e.g. describing bitrates and resolutions) are generated and stored at servers \cite{Stockhammer2011,Thang2010}. Based on the information of metadata as well as the terminal and networks, a client can adaptively decide which/when data parts should be downloaded. Currently, the question of video adaptation for HTTP streaming is still an open issue \cite{Begen2011_2}.

Various adaptation methods for HTTP streaming have been proposed over the past few years \cite{Liu2011,Miller2012,Zhou2014,Akhshabi2012,Muller2012,Thang2013,Duc2015,Hung2014,Evensen2011,Hung2013,Tian2012}. These methods can be roughly classified into two groups, throughput-based and buffer-based, each has its own strengths and weaknesses \cite{Thang2014}. Throughput-based methods decide the version based on the estimated throughput only, while buffer-based methods mainly use buffer characteristics as references for making decisions. Throughput-based methods are usually able to react quickly to throughput variations; however, the streaming quality may be unstable \cite{Liu2011}. Meanwhile, buffer-based methods try to maintain a smooth video stream, but may cause sudden changes in video quality when the buffer level drastically drops \cite{Miller2012,Zhou2014,Akhshabi2012,Muller2012}. It should be noted that as the datasize of each segment varies according to the requested version, the buffer size and buffer level should be measured in seconds of media.

So far, existing adaptation methods have mostly focused on CBR (constant bitrate) videos. The research on HTTP streaming for VBR (variable bitrate) videos is still limited. The problem with VBR videos is that, even though the throughput is stable, strong fluctuations of video bitrate may result in buffer underflows \cite{Thang2013}. Our previous work in \cite{Thang2013} is the first study on HTTP streaming that supports VBR video by estimating both the instant bitrate and the instant throughput. In the context of managed IPTV networks, where the bandwidth is allocated in advance, the delay-quality tradeoff of VBR video is optimally achieved by switching some high-bitrate segments \cite{Duc2015}. In \cite{Zhou2014}, a buffer-based adaptation method is proposed for VBR video streaming by using a partial-linear buffer prediction model along with a strategy to select versions in different buffer ranges. However, this method cannot avoid sudden changes of quality when the available bandwidth is drastically reduced.

In this paper, we present a novel adaptation method which can effectively support VBR videos. By extending our preliminary work in \cite{Hung2014}, the proposed method can cope with the variations of throughput as well as video bitrate. Especially, our method takes into account the moving average of bandwidth and video bitrate to provide stable streaming quality without sudden changes. The experimental results show that our approach can provide consistent VBR video streaming with smooth video quality and stable buffer level. To the best of our knowledge, this is the first method that can provide smooth version transitions for VBR video streaming over HTTP.

The organization of this paper is as follows. Section \ref{sec:1} presents an overview of HTTP streaming and the related work. The principles of our method as well as the algorithm description are presented in detail in Section \ref{sec:2}. Section \ref{sec:3} and \ref{sec:4} provide our experimental results and discussions, respectively. Finally, conclusion and direction for future work are given in Section~\ref{sec:concs}.

\section{Related Work}
\label{sec:1}
\begin{figure}
	\centering
	\includegraphics[width=\columnwidth]{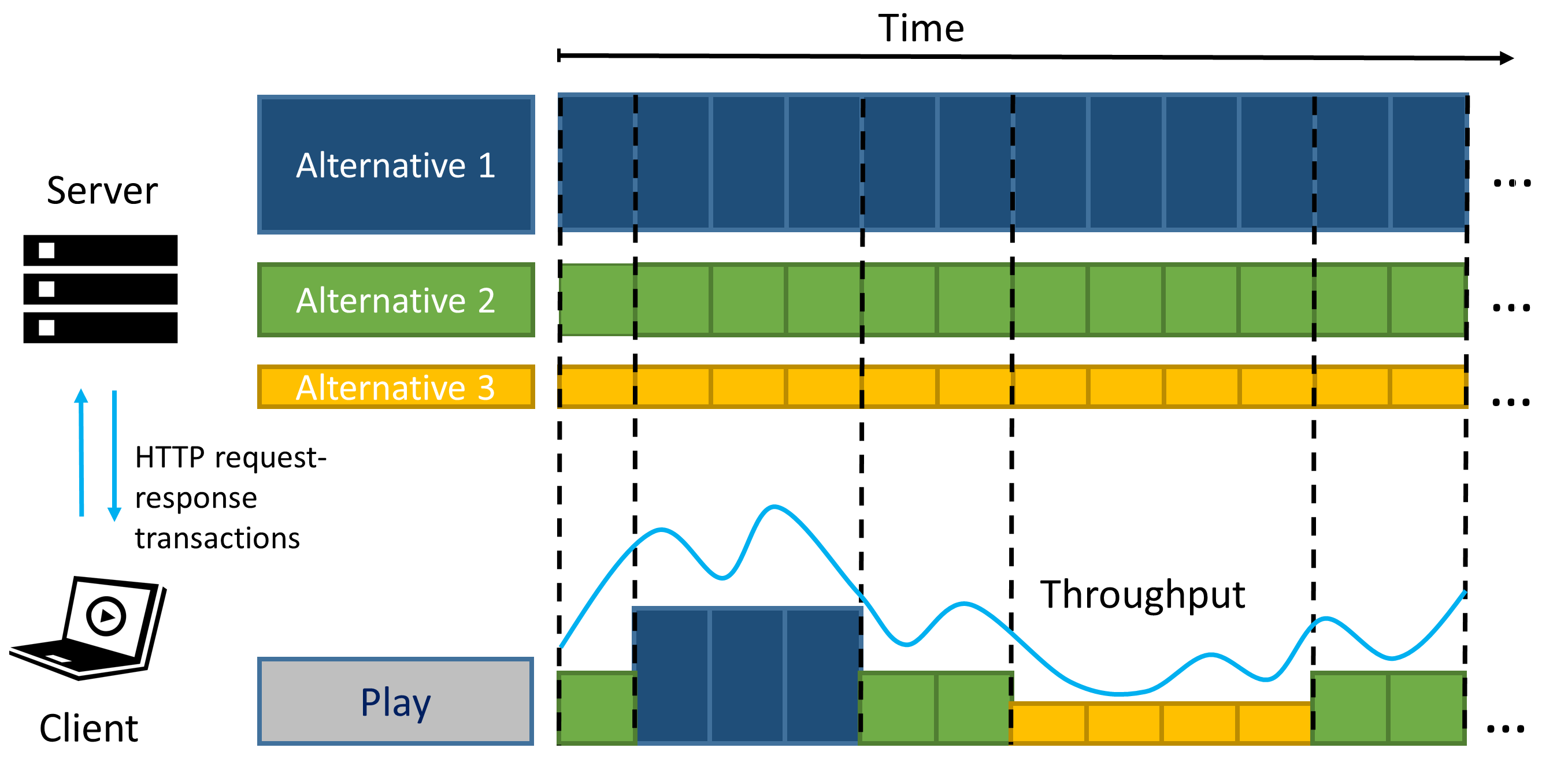}
	\caption{Media delivery hierarchy in HTTP streaming}
	\label{fig:hierarchy}
\end{figure}

The general architecture of an HTTP streaming system consists of servers, delivery networks and clients \cite{Thang2012,Stockhammer2011}. In MPEG DASH terminology, to support adaptivity a video is encoded in multiple versions (also called alternatives or representations), each of which is further divided into short segments. Video segments together with metadata are hosted at a server and will be requested by the client. In most cases, for each request from the client, the server will send one segment. Therefore, a video will be delivered by a sequence of HTTP request-response transactions. The version (low or high) of a requested segment is decided based on the metadata and status of terminals/networks. Fig.~\ref{fig:hierarchy} depicts an illustration of media delivery in DASH. More information about the HTTP streaming structure as well as DASH concepts could be found in \cite{Begen2011_1,Stockhammer2011}.

In general, an adaptation method needs to answer two key questions: 1) should the current version be maintained? and 2) if not, which version should be switched to? As already mentioned, existing methods can be divided into a throughput-based group and a buffer-based group. Throughput-based methods are different in the ways they estimate or use the throughput. In terms of throughput estimation, the simplest way is to use the measured throughput right after having fully received a segment (called instant throughput) as the throughput estimate of the next segment. Another approach is to use a smoothed throughput measure \cite{Thang2012}\cite{Akhshabi2012} to avoid short-term fluctuations, which are a drawback of using instant throughput. However, this may cause late reaction of the client to large throughput drops. In \cite{Thang2012} we propose a throughput estimation method that has the advantages of both instant throughput and smoothed throughput. Furthermore, other studies also present ways to obtain the estimated throughput based on sampled throughput values and RTT \cite{Thang2013}, probing or stored data (lookup table) \cite{Evensen2011}. Once the client obtains the estimated throughput, the version can be decided in many ways. A simple solution is using a safety margin to compute an appropriate version for the next segment \cite{Thang2012}. In \cite{Liu2011}, the version is controlled by a TCP-like mechanism where a measure proportional to the instant throughput is used as the key input.

Buffer-based methods, which mainly use buffer characteristics to decide the video versions, may take into account a throughput estimate as well. A popular strategy of these methods is dividing the buffer into multiple ranges with buffer thresholds $\beta_1,\beta_2,\beta_3,\beta_{max}$ $(0<\beta_1<\beta_2<\beta_3\le\beta_{max})$ \cite{Miller2012,Akhshabi2012,Muller2012}. When the buffer level stays in different ranges, different actions are applied. For instance, methods of \cite{Miller2012} and \cite{Akhshabi2012} try to maintain the current version in a specific buffer range (e.g. $\beta_2  \sim \beta_{max}$  for the method of \cite{Miller2012} and $\beta_2  \sim \beta_3$ for the method of \cite{Akhshabi2012}) and dynamically switch up/down the quality in other buffer ranges. Meanwhile, the method of \cite{Muller2012} chooses video versions following the variations of instant throughput (also with the use of up-scaling and down-scaling factors) based on different buffer ranges. In \cite{Hung2013} we introduce a trellis-based method that represents all possible changes of the versions and corresponding buffer levels in the near future. Thus, this approach can make good decisions on the bitrates of some future segments. In \cite{Tian2012}, the buffer level deviation and instant throughput are employed as inputs of a proportional-integral controller for adaptation.

\begin{figure}
	\centering
	\includegraphics[width=\columnwidth]{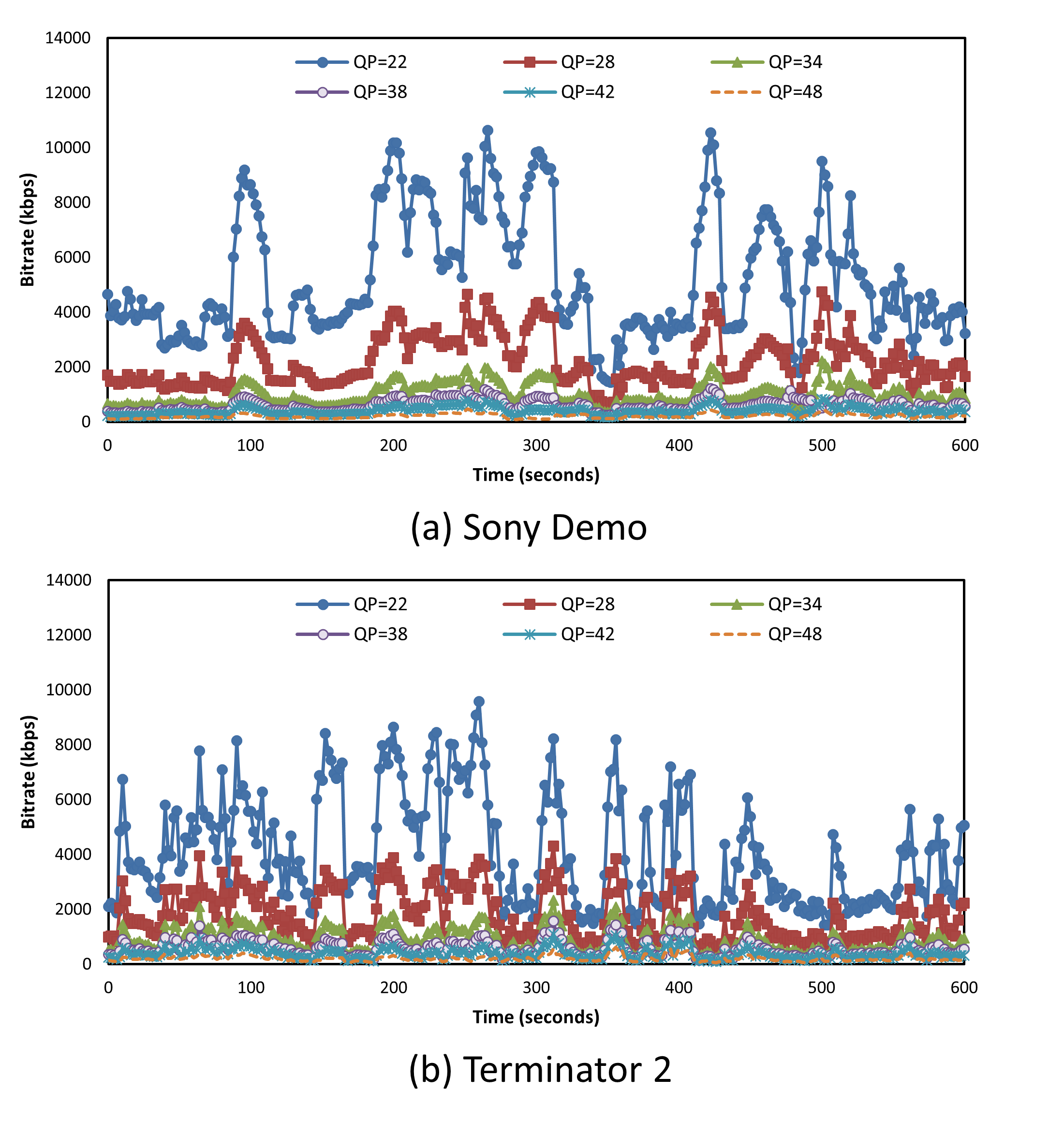}
	\caption{Bitrates of the versions of two test videos \cite{Trace2015}}
	\label{fig:testVideos}
\end{figure}

Yet, most of the existing methods have been developed only for the context of CBR videos, where the bitrate of a version is constant. Compared to CBR videos, videos encoded in VBR mode have important advantages in terms of quality and network resource usage \cite{Lakshman1998}. However, the variations of video bitrate over time, together with throughput fluctuations, result in a big challenge for HTTP adaptive streaming \cite{Thang2013}. Two examples of how bitrates of different versions vary, especially in some scene changes, are illustrated in Fig. \ref{fig:testVideos}. Detailed information of these videos will be described in Section~\ref{sec:3}. 

Our previous work in \cite{Thang2013} is the first study on VBR video streaming over HTTP. Besides throughput estimation, this method also considers the estimation of the instant video bitrate, which can be divided into 1) intra-stream estimation and 2) inter-stream estimation. The former estimates the bitrates of segments within a version, while the latter estimates the bitrates of segments across different versions. This method can provide a very stable buffer, and moreover, can support a CBR-like streaming service from VBR videos. In \cite{Zhou2014}, a buffer-based adaptation method for VBR videos is proposed, where the buffer is divided into multiple ranges. In order not to take into account varying video bitrates, this method uses a partial-linear trend prediction of the buffer level for choosing versions in different buffer ranges. If no significant change of buffer level is estimated, the client will maintain the current version for the next segment. However, this method still causes sudden version changes if the actual buffer level declines drastically. 

In this paper, we propose a novel adaptation method that can effectively support VBR videos in on-demand streaming. The distinguishing features of our method include:
\begin{itemize}
	\item To cope with strong variations of video bitrates, we propose using a local average bitrate as the (moving) representative bitrate of a version. Since this representative bitrate is stable in short term, it helps the client clearly differentiate the available versions and make a good version selection at each time instance. 
	\item Because the client does not have enough information about the segment bitrates of all versions, we provide an algorithm to obtain an estimated representative bitrate of each version.
	\item Regarding the first key question, quality stability is effectively maintained by 1) using a smoothed throughput estimate when the buffer is not in danger, 2) using the representative bitrate, and 3) being conservative in quality switching.
	\item Regarding the second key question, smooth transitions of quality are supported by avoiding jumping simply to the lowest version in panic case and by early switching down when there is a throughput-bitrate mismatch.
\end{itemize}
 
\section{The proposed method}
\label{sec:2}
In this section, we present a new buffer-based quality adaptation method for VBR video streaming. Some notations along with their definitions used in the paper are provided in Table~\ref{tab:notation}.

\begin{table}[ht]
\renewcommand{\arraystretch}{1.3}
\caption{Notations and definitions}
\centering
\label{tab:notation}      
	\begin{tabular}{c p{6cm}} 
	\hline
	\textbf{Notation} 		& \textbf{Definition}   \\[3pt]
	\hline
	$T_i$ 			& The throughput of segment $i$  \\
	$T_{i+1}^{est}$ & The throughput estimate of segment $i+1$  \\
	$\beta_{cur}$ 	& The current buffer level \\
	$\beta_{min}$ 	& The minimum buffer threshold \\
	$\beta_i^{th}$ 	& The flexible buffer threshold \\
	$\beta_{max}$ 	& The buffer size and also the target buffer level \\
	$B_{i,k}$ 		& The bitrate of the segment $i$ in version $k$. It could be the actual value $B_{i,k}^o$ or the estimated value $B_{i,k}^e$ \\
	$B_{i,k}^{rep}$ & The representative bitrate of version $k$ at segment $i$ \\
	$V$ 			& The number of available video versions \\
	$I_i$ 			& The index of the version which is chosen for segment $i$ (version of higher quality has a higher index value) \\
	\hline
	\end{tabular}
\end{table}

\subsection{Handling throughput and bitrate fluctuations}
Generally, based on the measured throughput and the video bitrate, the client should choose an appropriate version for the next segment. Suppose that, after receiving the current (or last) segment $i$ of version $I_i$, the client measures the bitrate $B_{i,I_i}$ and throughput $T_i$ of this segment. Now the client will decide the version $I_{i+1}$ for the next segment $i+1$. Our proposed method will also leverage the client buffer to cope with the fluctuations of both the throughput and the video bitrate. Depending on the current buffer level $\beta_{cur}$, the client will decide whether the version should be increased, decreased, or maintained. 

For the version selection of the next segment, it is necessary to estimate the throughput based on the throughput history of received segments. To avoid the effects of short-term fluctuations of instant throughput, we use a smoothed throughput measure $T_i^s$ \cite{Thang2012,Akhshabi2012} as the throughput estimate $T_{i+1}^{est}$ for the next segment $i+1$:
\begin{equation}
T_{i+1}^{est}=T_i^s=
	\begin{cases}
	(1-\delta) \times T_{i-1}^s + \delta \times T_i & \text{ if } i > 0 \\ 
	T_i & \text{ if } i = 0 
	\end{cases}
\end{equation}
where $\delta$ is a weighting value, which is set to $0.1$ in this paper. 

As video bitrate is highly fluctuating, we propose using a representative bitrate for each version, which can differentiate the available versions and can also be appropriate for maintaining quality stablity. Denote $B_{i,k}^{rep}$ the representative bitrate for version $k$ at segment index $i$. In our method, $B_{i,k}^{rep}$ is calculated as the average bitrate of $N$ recent segments of version $k$. 

The problem is that the client only knows the bitrates of the received segments, which may belong to different versions. So, after receiving each segment $i$, we will estimate the segment bitrates of other versions with the same index $i$. This is enabled by the bitrate estimation method proposed in our previous study \cite{Thang2013}, where the segment bitrates of other versions are estimated from the bitrate of the received segment using the inter-stream bitrate prediction. Specifically, the estimated bitrate $B_{i,k}^e$ of version $k$ can be calculated from the (actual) bitrate $B_{i,n}^o$ of the received segment $i$ with the selected version $n$ as follows:
\begin{equation}
	B_{i,k}^e=\theta \times B_{i,n}^o \times 2^{\frac{QP_n-QP_k}{6}}
	\label{eq:rateEst}
\end{equation}
where $QP_k$ and $QP_n$ are the quantization parameter (QP) values of the versions, and $\theta=1.05$ is an empirical factor used as the compensation for the approximation error of the model \cite{Thang2013}. In our notation (Table \ref{tab:notation}), bitrate $B_{i,k}$ can be either the actual bitrate $B_{i,k}^o$ or the estimated one $B_{i,k}^e$. Once obtaining the bitrates of all segments, the representative bitrate of each version at segment index $i$ is calculated as the average of the bitrates of segment $i$ and $N-1$ previous segments in that version. The algorithm to calculate the representative bitrates is provided in Algorithm \ref{alrm:rateEst}. In Section \ref{sec:3}, we will investigate how different values of $N$ affect the performance of our method.

\begin{algorithm}[ht]
	\renewcommand{\arraystretch}{1.3}
	\caption{Representative bitrate computation (after receiving the last segment $i$ and for all video versions)}
	\label{alrm:rateEst}
	\KwIn{$N,\ {B_{i-1,k}^{rep}|}_{1 \le k \le V},\ {B_{i-j,k}|}_{0 \le j < N, 1 \le k \le V}$}
	\KwOut{${B_{i,k}^{rep}|}_{1 \le k \le V}$}
	\For {$k \leftarrow 1,2,...,V$} {
		// Check if bitrate is the original bitrate \\
		\uIf {$k = I_i$} {
			$B_{i,k} \leftarrow B_{i,k}^o$; \\	
		} 
		// Otherwise the bitrate is the estimated bitrate \\
		\Else {
			Estimate $B_{i,k}^e$  by (\ref{eq:rateEst}); \\
			$B_{i,k} \leftarrow B_{i,k}^e$; \\
		}
		// Compute $B_{i,k}^{rep}$  \\
		$B_{i,k}^{rep} \leftarrow B_{i-1,k}^{rep}  + (B_{i,k}-B_{i-N,k})/N$;	            
	}
	
\end{algorithm}

\subsection{Adaptation algorithm}
Our algorithm will address the two key questions above, so as to avoid rebuffering and to reduce the number and the degree of (version) switches. Based on the current buffer level of the client, we define four possible cases in a streaming session, which are uptrend, stable, downtrend, and panic cases. In these four cases, the client will be likely to switch up, maintain, switch down, or aggressively decrease the version. For this purpose, our method divides the buffer into three ranges with thresholds $\beta_{min}$ and $\beta_i^{th}$ ($\beta_{min} < \beta_i^{th} < \beta_{max}$). Here $\beta_{max}$ is the buffer size and also the target buffer level of the adaptation method. 

When the current buffer level exceeds $\beta_{max}$, the uptrend case is activated. (For the reason why the buffer level could be higher than $\beta_{max}$, please refer to our previous work \cite{Thang2014}). However, it is not good if the client frequently goes back and forth between the uptrend case and downtrend case. To avoid this fluctuation, our method switches up the version by one version only if the representative bitrate $B_{i,I_i+1}^{rep}$ of the next higher version $I_i+1$ is smaller than the throughput estimate of the next segment (i.e. $B_{i,I_i+1}^{rep} <T_{i+1}^{est})$; otherwise, the client will maintain the current version. 

The stable case is determined by the condition $\beta_i^{th} \le \beta_{cur} < \beta_{max}$. In this case, the buffer level is judged as in a very safe condition, so no change in quality is needed. The client just maintains the current version to avoid unnecessary switches.

The downtrend case is activated when $\beta_{min} \le \beta_{cur} < \beta_i^{th}$. In this case, the client needs to carefully decide the requested version in order to avoid buffer underflows as well as sudden quality changes when the throughput and/or the video bitrate change drastically. In general, the version should be decreased; however, it is unnecessary to switch down always when the buffer level is in this range. In this process, we define the target bitrate for the next segment $i+1$ is the highest representative bitrate, which is lower than the throughput estimate: $B_{i+1}^{tar}= \max\{B_{i,k}^{rep}| B_{i,k}^{rep} < T_{i+1}^{est}\}$. If the instant bitrate and the representative bitrate do not exceed the target bitrate ($B_{i,I_i} \le B_{i+1}^{tar}$ and $B_{i,I_i}^{rep} \le B_{i+1}^{tar}$), the current version will be maintained. Otherwise, the client will switch down to the next lower version.

We can see that sometimes the instant throughput might be much smaller than the instant bitrate. Even though the buffer is not in danger yet, the downtrend case should be activated earlier to avoid having to quickly reduce the version in the future. This is enabled by increasing the value of $\beta_i^{th}$. In our method, $\beta_i^{th}$ is controlled using a logistic function as follows:
\begin{align}
&\beta_i^{th} = \beta_{max} - \frac{1}{1+e^\sigma} \times (\beta_{max} - \beta_{min}) \\
&\text{where } \sigma = 1 - \frac{T_i}{B_{i,I_i}}.
\end{align}
In this way, the larger the mismatch between $T_i$ and $B_{i,I_i}$ becomes, the higher the value of $\beta_i^{th}$ will be, while still satisfying the condition $\beta_{min}\le\beta_i^{th}<\beta_{max}$. 

The final case, called the panic case, is when the buffer level is in danger, i.e. $\beta_{cur}<\beta_{min}$. To ensure that the buffer will not be empty, the client will aggressively switch down the version. Yet, instead of jumping directly to the lowest version, the client adopts the method of our previous work \cite{Thang2013} in this case. Specifically, the client will choose a version, of which the instant bitrate is the highest but still lower than the instant throughput:
\begin{equation}
	I_{i+1} = \arg\!\max_{1 \le k \le V} \{B_{i,k} | B_{i,k} < T_i\}.
	\label{eq:panic}
\end{equation}
The algorithm of our method is summarized by pseudo code as in Algorithm \ref{alrm:bitrate}.

\begin{algorithm}[ht]
	\caption{Adaptation algorithm for VBR video streaming}
	\label{alrm:bitrate}
	\KwIn{$T_i,\ \beta_{cur}, \ I_i, \ B_{i,I_i}, \ B_{i,I_i}^{rep}, \ T_{i+1}^{est}$}
	\KwOut{$I_{i+1}$}
	// \textit{Uptrend case} \\
	\lnl{} \uIf {$\beta_{cur} > \beta_{max}$} {
		\uIf {$B_{i,I_i}^{rep} < T_{i+1}^{est}$} {
			$I_{i+1} \leftarrow I_i + 1$;
		} \Else {
			$I_{i+1} \leftarrow I_i$;
		}
	}
	// \textit{Stable case} \\ 
	\lnl{} \uElseIf{$\beta \in [\beta_i^{th},\beta_{max})$} {
		$I_{i+1} \leftarrow I_i$;
	} 
	// \textit{Downtrend case} \\
	\lnl{} \uElseIf{$\beta \in [\beta_{min},\beta_i^{th})$} {
		$B_{i+1}^{tar} \leftarrow \max \{ B_{i,k}^{rep}| B_{i,k}^{rep}<T_{i+1}^{est}\}$; \\
		\uIf {$B_{i,I_i} \le B_{i+1}^{tar} \text{ and } B_{i,I_i}^{rep} \le B_{i+1}^{tar}$} {
			$I_{i+1} \leftarrow I_i$;
		} \Else {
			$I_{i+1} \leftarrow I_i-1$;
		}	
	} 
	// \textit{Panic case} \\
	\lnl{} \Else{
		Select $I_{i+1}$ based on Eq. (\ref{eq:panic});
	}
	
\end{algorithm}

Generally, it can be seen that our method tries to provide a smooth video stream by two techniques. First, it is somewhat conservative in increasing the quality because that is allowed only when the buffer is full. In addition, the selected version is constrained by the long-term values of throughput and bitrate. Second, the downtrend case is also conservative by avoiding continuously switching down the quality. Meanwhile, in the panic case, we take into account the instant bitrate and instant throughput. Thus, the client can switch down gradually when possible; and there is no need to switch to the lowest version if the instant bitrate of a higher version still meets the throughput constraint.

\section{Experimental Results and Discussions}
\label{sec:3}
In this section, we will evaluate our proposed method and two reference methods in the context of on-demand VBR streaming, focusing on the behaviors of version switching and buffer level after the initial buffering stage. Two bandwidth traces, a simple one and a complex one, are employed in our experiments.

\subsection{Experiment setup}
\begin{figure}[t]
	\centering
	\includegraphics[width=\columnwidth]{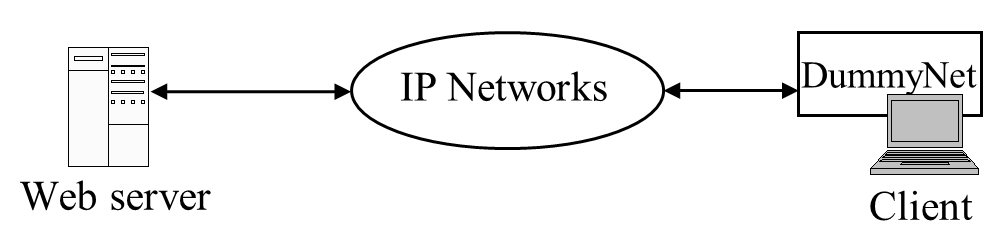}
	\caption{Test-bed organization for experiments}
	\label{fig:testbed}
\end{figure}

Our test-bed organization used for the experiments is similar to that of \cite{Thang2014}, which consists of an HTTP webserver, a streaming client and IP networks (Fig. \ref{fig:testbed}). The server is an Apache HTTP server of version 2.2.21 running on Ubuntu 12.04. Our test-bed uses DummyNet tool \cite{Rizzo1997} installed at the client side to emulate network characteristics. The packet loss rate is set to $0\%$, assuming that the bandwidth trace used in the experiments already takes into account the fluctuations caused by packet loss. RTT value of DummyNet is set to 40ms. The client is implemented in Java and runs on a Windows 7 notebook with Core i5 2.6GHz CPU and 4GB RAM.

The test videos are ``Sony Demo" and ``Terminator 2" \cite{Trace2015} with a frame rate of 30fps and a resolution of 1280x720. The duration of each video is 600 seconds. We encode 6 VBR versions by the High profile of H.264/AVC \cite{Wiegand2003}, corresponding to 6 different values of QP, namely 22, 28, 34, 38, 42 and 48. Each version is divided into small video segments of 2 seconds. The version index, QP, and the average bitrate of each version are listed in Table \ref{tab:videoInfo}. The bitrate traces of the video versions are shown in Fig. \ref{fig:testVideos}. 

\begin{table}[ht]
	\renewcommand{\arraystretch}{1.3}
	\caption{Version information of the two test videos}
	\centering
	\label{tab:videoInfo}      
	\begin{tabular}{c c c c c} 
		\hline
		\bfseries Index	& \bfseries	QP	& \multicolumn{2}{c} {\bfseries	Average bitrate (kbps)} \\
		\cline{3-4}
			&		&	Sony Demo	&	Terminator 2 \\
		\hline
		1	&	48	&	203.77	&	201.55	\\
		2	&	42	&	390.75	&	377.97	\\
		3	&	38	&	602.96	&	567.02	\\
		4	&	34	&	991.32	&	882.29	\\
		5	&	28	&	2194.05	&	1798.93	\\
		6	&	22	&	5180.58	&	4127.86	\\
		\hline
	\end{tabular}
\end{table}

As we focus on on-demand streaming, the buffer size of the client is set to 50s (i.e. 25 segment durations). For comparison, the two reference methods which are the instant throughput - instant bitrate based method \cite{Thang2013} (called ITB) and the buffer-based method with trend prediction \cite{Zhou2014} (called TBB) are implemented. The TBB method is implemented with buffer thresholds ($B_{min}, B_{low}, B_{high}, B_{max}) = (10s, 20s, 40s, 50s)$. In our method, $\beta_{min}$ and $\beta_{max}$ are 10s and 50s, respectively. Also, we investigate whether the number of segments $N$ used for representative bitrates affects our method's performance. The values of $N$ being considered are 10, 30 and 50, and these options of our method are referred to as AVG-10, AVG-30 and AVG-50, respectively.

\subsection{Simple bandwidth scenario}
\begin{figure}[t]
	\centering
	\includegraphics[width=0.75\columnwidth]{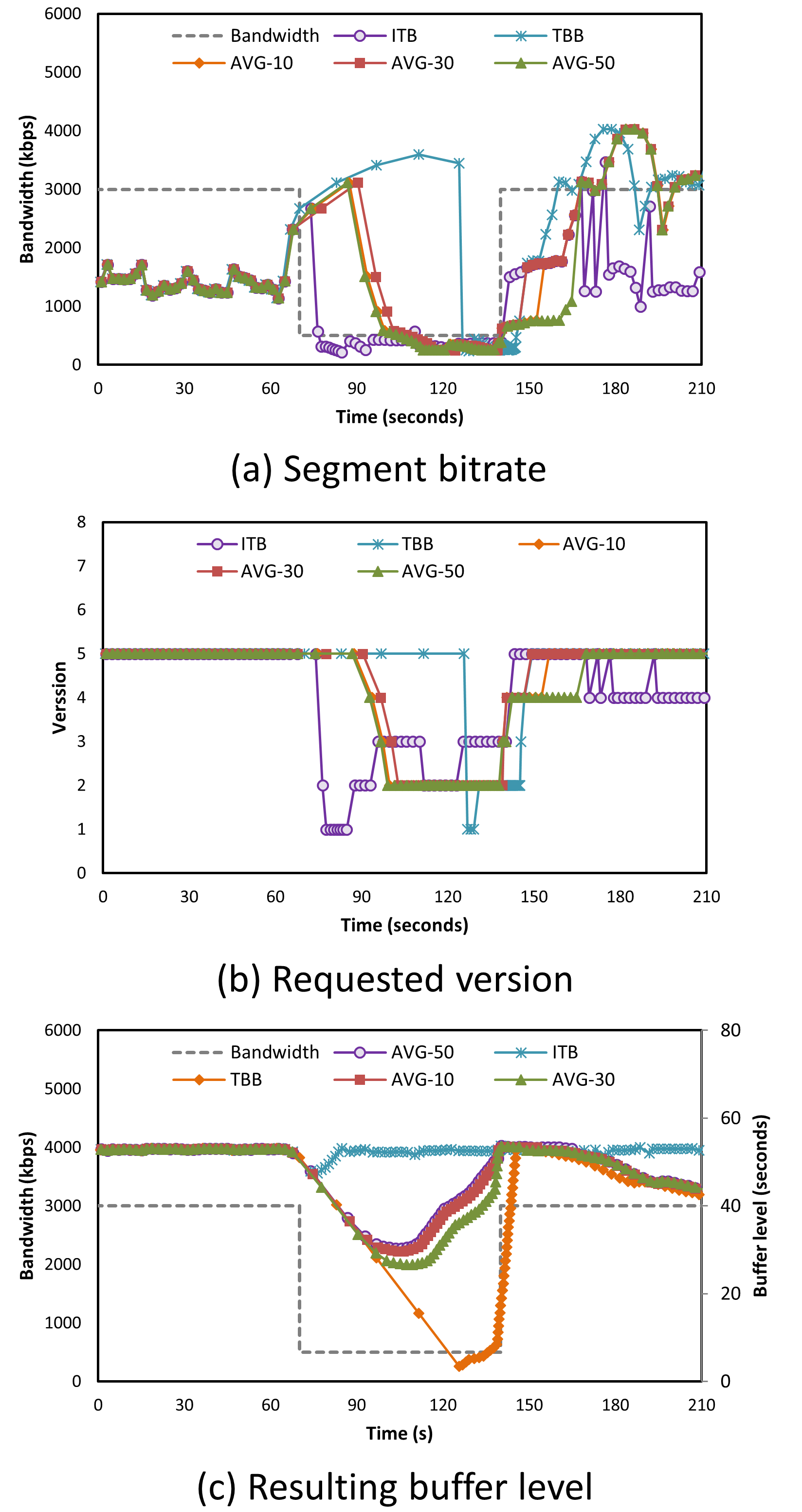}
	\caption{Adaptation results of the three methods in simple bandwidth scenario}
	\label{fig:simple}
\end{figure}

First, we investigate the performance of the methods in a simple bandwidth scenario, when the available bandwidth drops suddenly. The bandwidth has a rectangular shape with two bandwidth levels, 2500 kbps and 500 kbps, as shown in Fig. \ref{fig:simple}a. This case is important in evaluating adaptation methods because we need to know how they perform when the bandwidth drops drastically. In this case, the Sony Demo video is used.

Fig. \ref{fig:simple} shows the comparisons of requested versions, bitrates and buffer levels of the three methods. It is clear that the TBB method tries to maintain the high quality (version 5) for too long even when the available bandwidth drops and stays at the low level for a long interval, resulting in the worst buffer level curve (Fig. \ref{fig:simple}c) as well as a drastic drop of quality (around $t = 125s$, from version 5 to version 1 in Fig. \ref{fig:simple}b). As for the ITB method, because it requests versions following the variations of instant throughput and instant video bitrate, the quality is aggressively changed over time while the buffer level variations are very small.

As for our method, all the three options have similar behaviors in terms of bitrate, version switch, and buffer level. Our method reduces the video quality gradually with no switches larger than 1 while the buffer level is higher than 25 seconds. Also, the minimum version provided by our method is 2, while the other two methods sometimes jump to version 1. 

\subsection{Complex bandwidth scenario}
\begin{figure}[ht]
	\centering
	\includegraphics[width=\columnwidth]{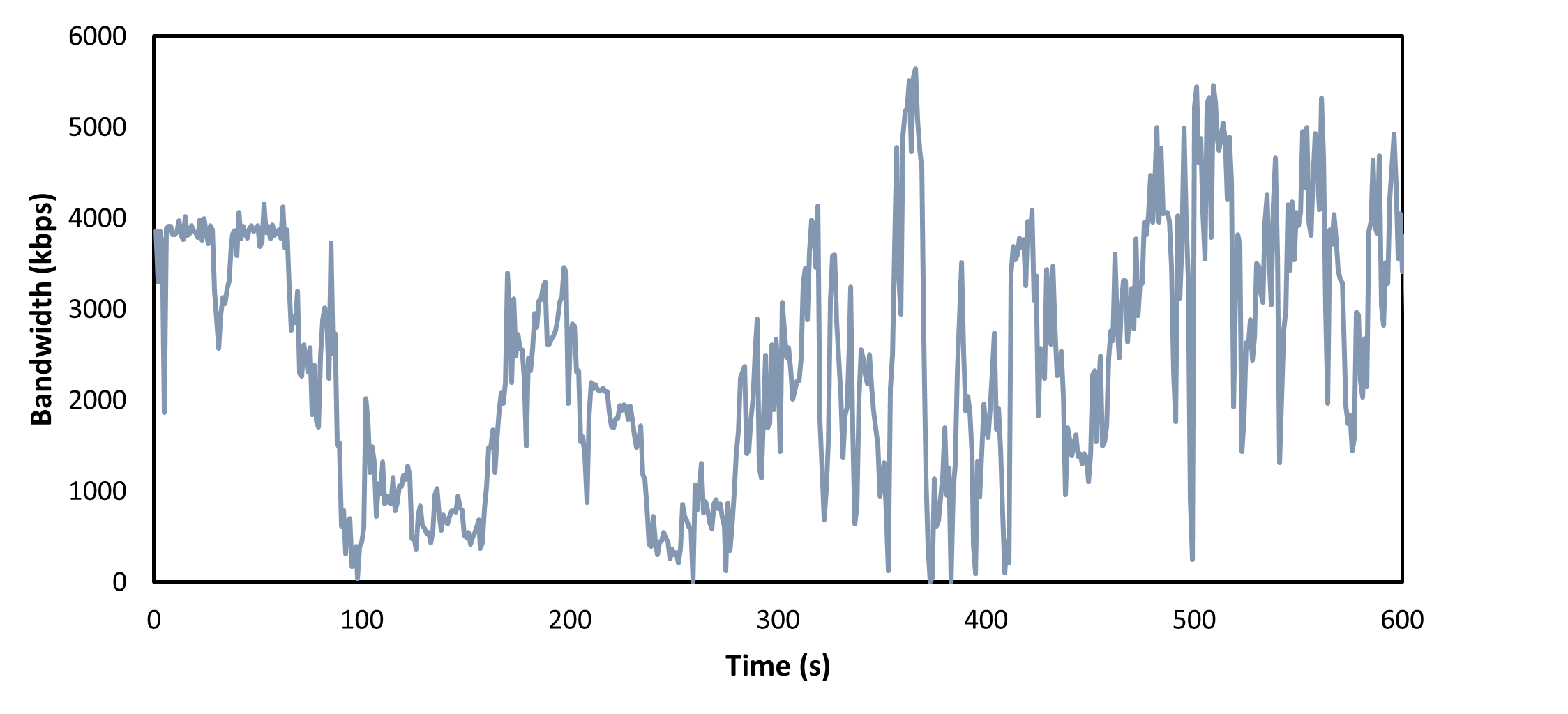}
	\caption{The bandwidth trace used in the complex bandwidth scenario}
	\label{fig:bandwidth}
\end{figure}

In this part, we evaluate the adaptation methods with a complex bandwidth trace (Fig. \ref{fig:bandwidth}), which was obtained from a mobile network \cite{Muller2012}. Both test videos, ``Sony Demo" and ``Terminator 2", are employed in this scenario.

\begin{figure}[ht]
	\centering
	\includegraphics[width=\columnwidth]{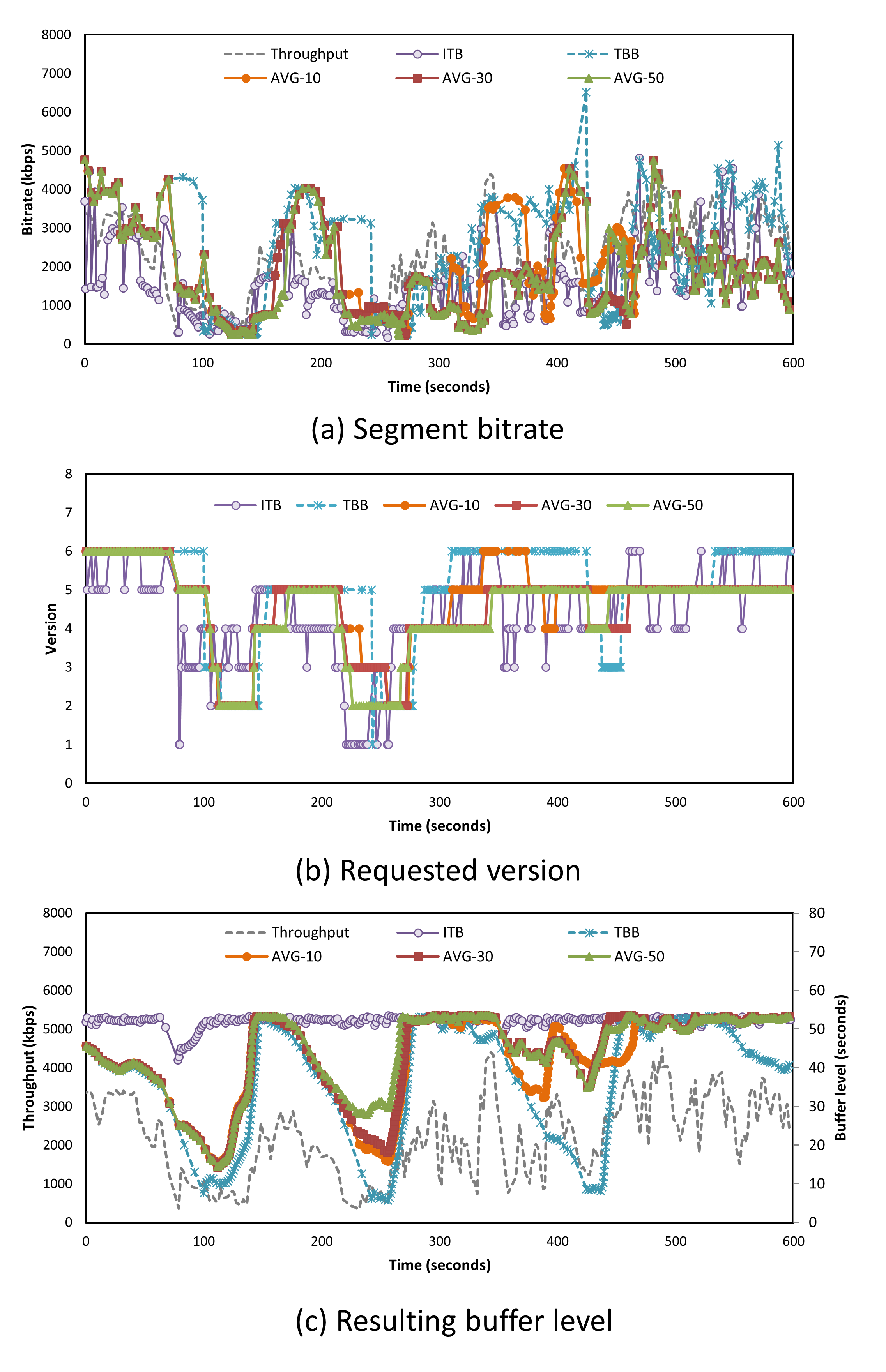}
	\caption{Adaptation results of the three methods in the complex bandwidth scenario with ``Sony Demo" video}
	\label{fig:sony}
\end{figure}

Fig. \ref{fig:sony} shows the experimental results for the ``Sony Demo" video. It is obvious that the TBB method's behavior is similar to that in the simple bandwidth case. Usually, this method tries to maintain a high version as long as the buffer level allows. This behavior results in sudden drops of quality (e.g. from version 6 to version 3 at 100s, and from version 5 to version 1 at 240s). Also, the buffer level curve of this method is the worst, as in the previous scenario. 

As for the ITB method, its bitrate curve closely follows the throughput curve, resulting in a highly fluctuating version curve with many switches, including switches of large degrees. Anyway, this method has the most stable buffer level curve as seen in Fig. \ref{fig:sony}c.

Meanwhile, our method provides both quality stability and buffer stability. The version curves of our method (Fig. \ref{fig:sony}b) have no sudden switches. Moreover, when the throughput decreases, the selected version of our method is always higher than or equal to those of the other methods. The buffer level curve of our method is not as stable as that of the ITB method; however, it is always higher than that of the TBB method. Compared to the TBB method, our method is more conservative in switching-up and less conservative in switching-down. 

Some statistics of the adaptation results are provided in Table \ref{tab:sonySta}. The statistics are related to bitrate, requested version, version switches, and buffer level. In terms of bitrate, the TBB method has the highest average bitrate since this method tends to select and maintain the best possible quality. However, its average version is interestingly lower than that of our AVG-10 option (4.19 vs. 4.30). In fact, the average version values of all methods are very similar (about 4.2 or 4.3). Among the three options, the AVG-10 option is the most aggressive as it always tries to request higher versions; howerver, this causes a little more switches than other options. These points will be explained and discussed further in the next part.

\begin{table}[ht]
	\renewcommand{\arraystretch}{1.3}
	\caption{Statistics of the reference methods of the proposed method with three different settings for “Sony Demo”. Here STD is the standard deviation}
	\centering
	\label{tab:sonySta}      
	\begin{tabularx}{\columnwidth}{l X X X X X} 
		\hline
		\textbf{Statistics}	& \textbf{ITB}	& \textbf{TBB}	&	\textbf{AVG-10}	&	\textbf{AVG-30}	&	\textbf{AVG-50}	\\
		\hline
		Average bitrate (kbps)	&	1555.2	&	1951.5	&	1777.7	&	1632.0	&	1637.1	\\
		\hline
		Average version			&	4.29	&	4.19	&	4.30	&	4.18	&	4.16	\\
		Maximum version			&	6		&	6		&	6		&	6		&	6		\\
		Minimum version 		&	1		&	1		&	2		&	2		&	2		\\
		\hline
		Number of switches		&	94		&	18		&	17		&	15		&	15		\\
		Maximum switch degree	&	4		&	4		&	1		&	1		&	1		\\
		STD of switch degrees	&	0.56	&	0.37	&	0.23	&	0.22	&	0.22	\\
		\hline
		Minimum buffer level (s)&	42		&	5.2		&	15.1	&	14.3	&	14.4	\\
		STD of buffer levels (s)&	1.59	&	15.31	&	11.49	&	11.41	&	10.43	\\
		\hline
	\end{tabularx}
\end{table}

\begin{figure}[ht]
	\centering
	\includegraphics[width=\columnwidth]{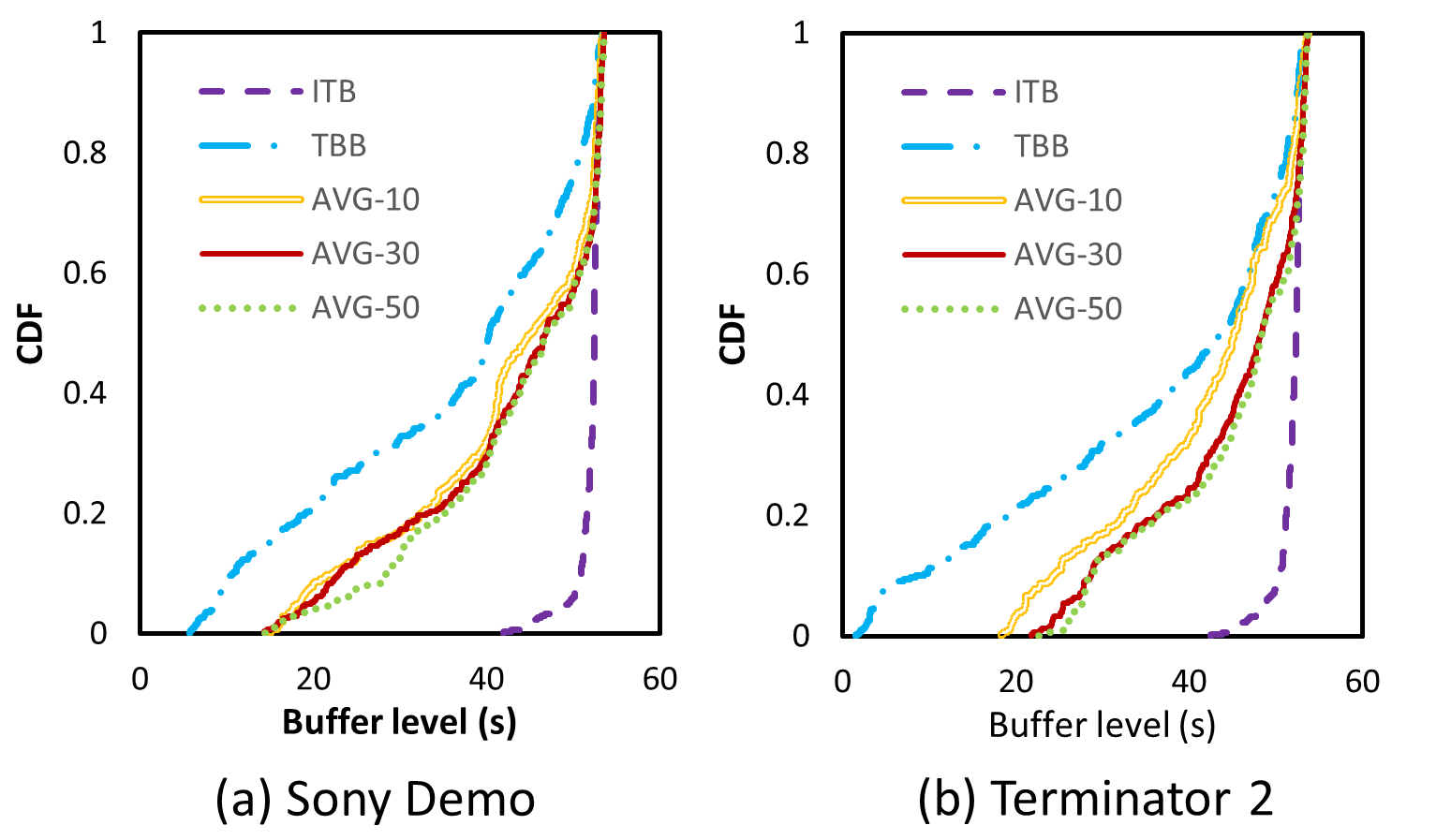}
	\caption{Cumulative distribution functions (CDF) of buffer level in the complex bandwidth scenario}
	\label{fig:cdf}
\end{figure}

Moreover, our method provides the smallest values in terms of the number of switches and the degree of switches. In addition, the minimum version of our method is higher than the other methods. The minimum value and standard deviation (STD) of buffer level of our method are also much better than those of the TBB method. For more information about the buffer, the cumulative distribution functions (CDF) of the buffer levels are provided in Fig. \ref{fig:cdf}a. Again, it is evident that the buffer of our method is more stable than that of the TBB method.

\begin{figure}[ht]
	\centering
	\includegraphics[width=\columnwidth]{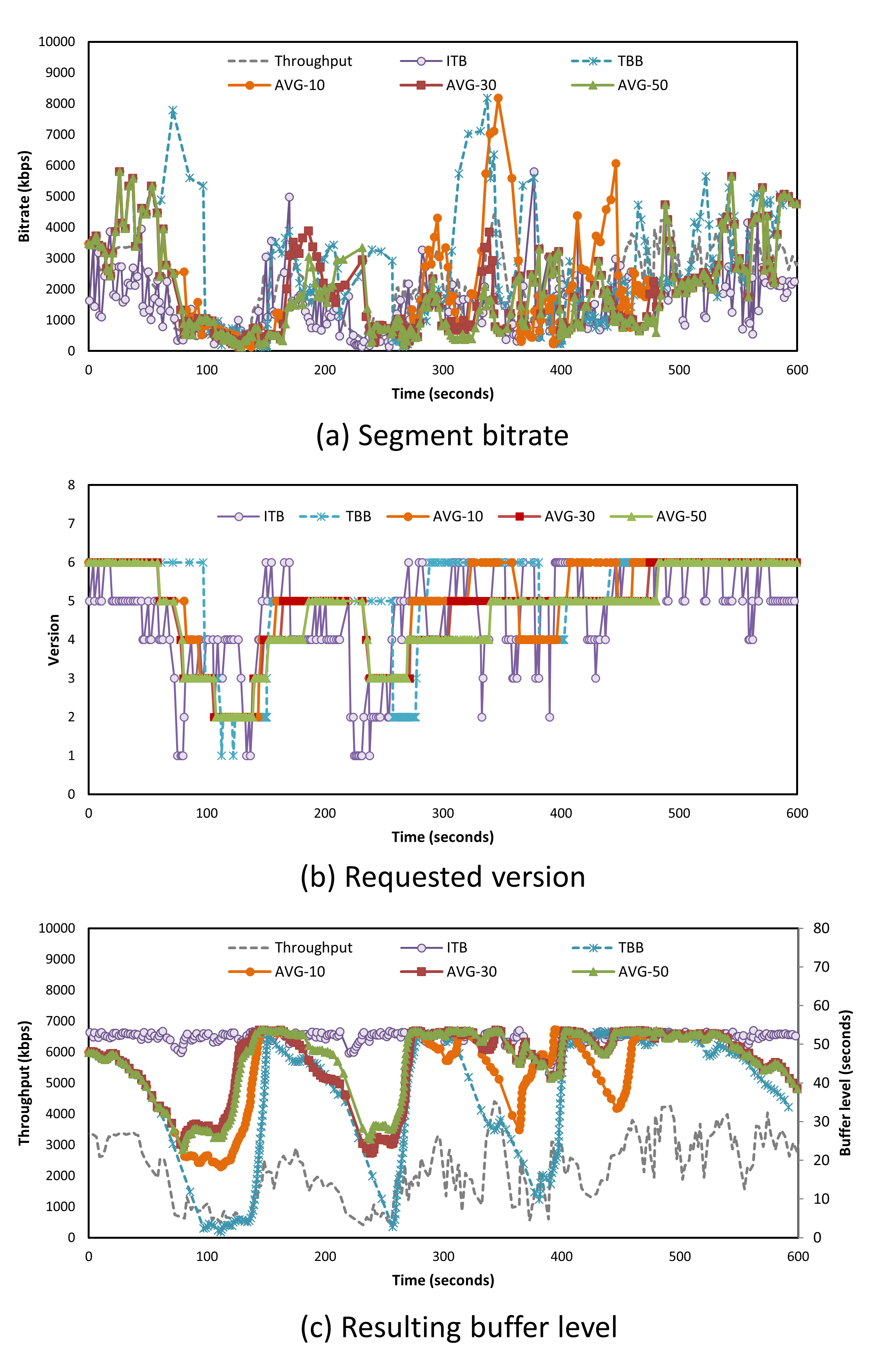}
	\caption{Adaptation results of the three methods in the complex bandwidth scenario with ``Terminator 2" video}
	\label{fig:terminator}
\end{figure}

Fig. \ref{fig:terminator} shows the experimental results for Terminator 2 video, where similar behaviors of the methods can be found. Our method still provides a smoother version curve and a more stable buffer level curve than the TBB method. The aggressiveness of AVG-10 can also be seen in Fig. \ref{fig:terminator}a. For example, during $330s \sim 370s$, the AVG-10 option shortly increases the quality to version 6 and then reduces the quality to version 5 and version 4. Meanwhile, the other two options still request version 5 during the same interval.

The statistics of the adaptation results for Terminator 2 video is provided in Table \ref{tab:terSta}. With this video, the average version values of our method are all higher than that of the TBB method. Especially, the average bitrate of the AVG-10 option is higher than that of the TBB method. This is because version bitrates of Terminator 2 are smaller than those of Sony Demo, so the AVG-10 option can easily reach the highest possible quality. In addition, the other parameters also show that our method provides a smoother quality and a more stable buffer that the TBB method. 

\begin{table}[ht]
	\renewcommand{\arraystretch}{1.3}
	\caption{Statistics of the reference methods of the proposed method with three different settings for ``Terminator 2". Here STD is the standard deviation}
	\centering
	\label{tab:terSta}      
	\begin{tabularx}{\columnwidth}{l X X X X X} 
		\hline
		\textbf{Statistics}	& \textbf{ITB}	& \textbf{TBB}	&	\textbf{AVG-10}	&	\textbf{AVG-30}	&	\textbf{AVG-50}	\\
		\hline
		Average bitrate (kbps)	&	1544.3	&	1926.2	&	1962.9	&	1779.4	&	1691.3	\\
		\hline
		Average version			&	4.51	&	4.34	&	4.61	&	4.55	&	4.46	\\
		Maximum version			&	6		&	6		&	6		&	6		&	6		\\
		Minimum version 		&	1		&	1		&	2		&	2		&	2		\\
		\hline
		Number of switches		&	128		&	17		&	18		&	12		&	12		\\
		Maximum switch degree	&	3		&	3		&	1		&	1		&	1		\\
		STD of switch degrees	&	0.63	&	0.34	&	0.24	&	0.20	&	0.20	\\
		\hline
		Minimum buffer level (s)&	42.5	&	1.5		&	18.3	&	21.8	&	22.7	\\
		STD of buffer levels (s)&	1.64	&	16.44	&	10.43	&	9.22	&	8.98	\\
		\hline
	\end{tabularx}
\end{table}

Besides, the CDFs of buffer levels are also (Fig. \ref{fig:cdf}b) confirm the stability of buffer level of our method. Fig. \ref{fig:cdf}b also shows that the buffer levels of the AVG-30 and AVG-50 options are a little better than that of the AVG-10 option. This is because the AVG-10 option is more aggressive than the other options. Especially, compared to the case of Sony Demo video (Fig. \ref{fig:cdf}a), the buffer of the TBB method is worse while the buffer of our method is better. This is actually due to the characteristics of the Terminator 2 video as discussed in the next part.

\section{Discussions}
\label{sec:4}
From the above experimental results, we can see some interesting points. First, sometimes the average bitrate of the TBB method is higher than that of our proposed method, while its  average version is lower than that of our method. This can be explained by the fact that sometimes the instant bitrate is very high. So, having some more segments of high bitrates, especially those belonging to the top version, will signficantly increase the average bitrate. In case of Terminator 2 video, which has lower version bitrates than Sony Demo video, the average bitrate of TBB method is not better than that of our method.

Meanwhile, in low-throughput periods, the selected versions of the TBB method are lower than those of our method, resulting in a lower value of average version. So, as the bitrate of VBR video is highly fluctuating, the parameter of average bitrate should not be the main indicator to evaluate the performance of VBR video streaming; rather, the parameter of average version should be considered. This is different from CBR video streaming, where the  average bitrate is always of high interest. 

The results of our adaptation method show that the use of a representative bitrate for each version is important for VBR videos. Even the ITB method benefits a lot from the simple estimation of the instant bitrate. Though having a highly fluctuating version curve, this method can be used for live streaming thanks to its very stable buffer. On the other hand, the TBB method does not consider bitrate values, and thus poorly handles the strong variations of VBR video. For example,  although Terminator 2 video has lower version bitrates than Sony Demo video, the TBB method's performance in the case of Terminator 2 video is much worse than in the case of Sony Demo video. This is just because the version bitrates of Terminator 2 video are extremely varying (with repeated and quick changes from a high value to a low value). Whereas, the performance of our method is not degraded in the case of Terminator 2 video.  
 
The three options of our method are similar in terms of average version values. Yet, among the three options, the AVG-10 option seems to be more aggressive in switching up and maintaining a high quality. This is because the representative bitrate of this option is more ``local" and can quickly detect low-bitrate segment intervals, where the client can switch to a higher version when the throughput allows. Such property can increase the average version; however, that also causes more switches as seen in the statistics of the AVG-10 option. The above statistics show that AVG-30 and AVG-50 options are nearly the same with good quality stability and good buffer stability. 

In general, it is shown that our method is more effective than the reference methods in providing stable streaming quality, with smooth version transitions even when the available bandwidth dramatically drops. If a user prefers streaming with the highest possible quality, more ``local" representative bitrates should be used. On the other hand, if the user prefers a stable streaming session with less version switches quality, more “global” representative bitrates should be used.

\section{Conclusion}
\label{sec:concs}
In this paper, we have presented an adaptation method for VBR video streaming over HTTP. To cope with strong variations of video bitrate, we employed a local average bitrate as the representative bitrate of a version. A buffer-based algorithm was then proposed to conservatively adapt video quality by taking into account a smoothed throughput estimate, the representative bitrate, and even the instant bitrate. The experimental results showed that our method can provide smooth video quality and stable buffer level, even under very strong variations of bandwidth and video bitrates. For future work, we will focus on live streaming scenarios such as surveillances with VBR video sources.

\bibliography{ref}

\end{document}